\documentclass[12pt]{article}
\begin{document}
\baselineskip=16pt

\title
{\begin{Huge}
Is There Unification in the 21st Century ?\\
\end{Huge}
\vspace{0.8in} }
\author{Yuan K. Ha\\ Department of Physics, Temple University\\
 Philadelphia, Pennsylvania 19122 U.S.A. \\
 yuanha@temple.edu \\    \vspace{.1in}  }
\date{July 1, 2010}
\maketitle
\vspace{.4in}
\begin{abstract}
\noindent
In the last 100 years, the most important equations in physics are Maxwell's equations for electrodynamics,
Einstein's equation for gravity, Dirac's equation for the electron and Yang-Mills equation for elementary 
particles. Do these equations follow a common principle and come from a single theory? Despite intensive 
efforts to unify gravity and the particle interactions in the last 30 years, the goal is still to be achieved.
Recent theories have not answered any question in physics. We examine the issues involved in this long quest
to understand the ultimate nature of spacetime and matter.\\
\end{abstract}

\newpage
\noindent
{\bf 1 \hspace{.1in} Introduction}\\

\vspace{.2in}
\noindent
At the end of the 19th century, physicists were very confident that they had the laws of nature at hand.
Classical mechanics had been firmly established for 200 years. Celestial mechanics was highly developed.
Electrodynamics was discovered. Thermodynamics was understood. The euphoria was so evident that in a speech
given by Albert Michelson in 1894 the following remarks were said [1]:
\begin{quotation}
\noindent
``{\em The more important fundamental laws and facts of physical science have all been discovered, and these are
so firmly established that the possibility of their ever being supplanted in consequence of new discoveries
is exceedingly remote ... Our future discoveries must be looked for in the sixth place of decimals.}''
\end{quotation}
Since then, physicists have identified four fundamental interactions together with their interaction strengths:\\

\noindent
1. Electromagnetic interactions - $10^{-2}$.\\
2. Weak interactions - $10^{-5}$.\\
3. Strong interactions - $10^{0}$.\\
4. Gravitational interactions - $10^{-38}$.\\

\noindent
They have also discovered a set of fundamental particles:
the quarks $(u, d, s, c, b, t)$;
the leptons $(e, \mu, \tau, \nu_{e}, \nu_{\mu}, \nu_{\tau})$; 
the gauge bosons $(\gamma, W^{+}, W^{-}, Z^{0}, G^{\alpha \beta})$, and the still anticipated 
Higgs boson $H^{0}$. These interactions and particles are governed by four fundamental equations:\\

\noindent
1. Maxwell's equations (1864).\\
2. Einstein's equation (1915).\\
3. Dirac's equation (1928).\\
4. Yang-Mills equation (1954).\\

\vspace{.3in}
\noindent
The theories constructed for the fundamental interactions are gauge theories based on various Lie groups:\\

\noindent
1. Quantum Electrodynamics - $U(1)$.\\
2. Quantum Electroweak Theory - $SU(2) \times U(1)$.\\
3. Quantum Chromodynamics - $SU(3)$.\\
4. Classical General Relativity - $SO(3,1)$.\\

\noindent
The goal of unification is not simply to combine the various fundamental interactions in a consistent
mathematical framework. It should entail a unifying principle and produce an interlocking structure.
It should answer long-standing questions and make new predictions. The search for unification would
force physicists to confront fundamental issues, to abandon old dogmas and to recognize new realities.
The quantum theory of elementary particles has been quite successful in that it can explain accurately
a number of phenomena in the strong, weak and electromagnetic interactions. On the other hand, it is
gravity which is the most challenging and the least understood of the four interactions. We shall
therefore focus our attention in this article to the problems in gravity. Without a deeper understanding
of the nature of gravity and the theories which claim to explain it, unification is pointless as recent
attempts to unify the interactions have not answered any question in physics.\\

\vspace{.2in}
\noindent
{\bf 2 \hspace{.1in} Supersymmetry}\\

\noindent
The goal of supersymmetry is to unify spacetime and internal symmetries of elementary particles, thereby
evading the Coleman-Mandula theorem which states that all possible symmetries of the S-matrix under general
assumptions can only be a direct product of the Poincare algebra and an internal symmetry algebra.
In supersymmetry, there exists a symmetry between fermions and bosons and the prediction is for every boson
there exists a corresponding fermion of the same mass and quantum numbers. The role of supersymmetry is to
cancel divergences in the perturbative calculations of quantum field theory since fermion and boson have
opposite signs in loop corrections. In the standard model, quadratic corrections to the Higgs mass due to
Yukawa interactions appear that cause its mass to diverge. Supersymmetry does away with the corrections by
supplying terms with a minus sign. This scheme works through some physical cutoff mechanism, and there is 
a scale associated with it. Within the dimensional regularization approach, however, quadratic divergences
do not exist and it is not clear what purpose would be served by a supersymmetric theory. From another point
of view, the goal of supersymmetry is not to double the number of fundamental particles. The doubling of 
particles has already been achieved by the existence of antiparticles. Antiparticles are crucial in virtual
particle pair creations and annihilations in quantum field theory. So far there is no irrefutable evidence
that supersymmetry is a symmetry of nature after 40 years. According to Veltman [2]:
\begin{quotation}
\noindent
``{\em The concept of naturalness is usually cited as the underlying motivation for supersymmetry. We will
challenge that concept, and in any case need to point out that there is nothing natural about the development
of the theory itself. Its main success is its agility in dodging the facts. The dubious explanation of the
convergence of the three scale coupling constants into a single point can not be taken seriously. It is just
another fit, using some of the many free parameters.}''
\end{quotation}
\noindent
It should be pointed out that coupling constant unification does not prove unification of the strong, weak and
electromagnetic interactions. There are other particles that can produce coupling constant unification.\\

\vspace{.2in}
\noindent
{\bf 3 \hspace{.1in} Higher Dimensions}\\

\noindent
Many unification theories involve higher spacetime dimensions. There is nothing compelling about higher dimensions
themselves. They may simply be a book-keeping device to account for the number of observed gauge fields. Gauge
transformations are coordinate transformations in higher-dimensional space. In general, higher-dimensional theories
suffer from instability and causality problems. There are negative energy solutions of the field equation. 
In Kaluza-Klein type theories of pure gravity in higher dimensions, the difficulty is noticeable at the classical
level. Analysis of the perihelion shift of planets in the solar system shows that the shift depends on the total
number of spatial dimensions in these theories [3]. The decomposition of metric assumes only a compact internal
space with the geometry of tori. The result is independent of the size of the extra dimensions, even if it is of
sub-millimeter scale. Starting from the multidimensional Einstein equation, a nonrelativistic limit of the metric
in four dimensions can be obtained. The metric coefficients are found to depend explicitly on the total number of
spatial dimensions $D$ and they affect the equation of motion in general relativity. In the perihelion shift 
calculation of the planet Mercury, the resulting formula is given by\\
\begin{equation}
\frac{D}{D-2} \left( \frac{\pi m^{2}c^{2}R^{2}_{S}}{2M^{2}} \right),
\vspace{.1in}
\end{equation}
where $m$ is the mass of the planet; $M$, the mass of the Sun; $R_{S}$, the Schwarzschild radius of the Sun, and
$c$ is the speed of light. The observed discrepancy for Mercury is $43.11 \pm 0.21$ arcsec per century. Only the ordinary
three-dimensional case $D=3$ gives a satisfactory result $42.94 ^{\prime\prime}$ which is within the measurement accuracy. For
$D=4$, the result is $28.63 ^{\prime\prime}$ and for $D=9$ it is $18.40 ^{\prime\prime}$. Thus all multidimensional case $D>3$ contradict observations.\\

In the deflection of light by the Sun, a corresponding analysis provides the formula [4]
\begin{equation}
\frac{D-1}{D-2} \left( \frac{R_{S}}{R} \right),
\vspace{.1in}
\end{equation}
where $R$ is the radius of the Sun. The observed deflection of a light ray that grazes the Sun surface has a
historical value of $1.75$ arcsec. For the three-dimensional case $D=3$ the above formula reproduces this value
accurately. For $D=4$, the result is $1.31 ^{\prime\prime}$ and for $D=9$ it is $1.00 ^{\prime\prime}$. Again, the multidimensional case shows
a severe problem with the classical tests of general relativity. The implication of incorporating Kaluza-Klein
type theories in unification is rather obvious.\\

\vspace{.2in}
\noindent
{\bf 4 \hspace{.1in} Higher Derivative Gravity Theories}\\

\noindent
A number of theories known as higher derivative gravity theories have the goal of constructing a renormalizable
theory of gravity explicitly in four dimensions. The Lagrangians contain higher order curvature invariants in
Riemannian geometry such as those of scalar curvature $R^{2}$, Ricci curvature $R_{\alpha \beta}R^{\alpha \beta}$,
Riemann curvature $R_{\alpha \beta \mu \nu}R^{\alpha \beta \mu \nu}$ and other combinations of these terms,
including Weyl curvature invariant $C_{\alpha \beta \mu \nu}C^{\alpha \beta \mu \nu}$, in order that the equations
be invariant under general coordinate transformations. These theories generally have problems with stability,
unitarity, ghosts and nonlocality [5]. None of them is yet successful as a quantum theory of gravity. A further
problem of higher derivative theories at the classical level is that none admits Birkhoff's theorem [6], which
states that spherically symmetric solution is unique and time-independent. The failure of Birkhoff's theorem in
higher derivative gravity theories means that spherically symmetric solution is time-dependent and dynamical.
A similar failure of Birkhoff's theorem in a generalization of Einstein's gravity called $f(R)$ theory, in which
the action is a nonlinear function of the scalar curvature $R$, also shows that spherically symmetric solutions
are time-dependent [7]. As a result, black holes in these theories are dynamical. Their horizons disappear and
a naked singularity will emerge [8]. In some $f(R)$ models, relativistic stars cannot exist due to the dynamics of
the effective scalar degree of freedom and there are doubts about the viability of these models [9].\\

\vspace{.2in}
\noindent
{\bf 5 \hspace{.1in} Alternative Gravity Theories}\\

\noindent
There are still other efforts to modify Einstein's gravity theory in order to achieve a finite and consistent theory
of quantum gravity. These are generally known as modified gravity theories [10]. The modification can take place both
at the microscopic scale and at the macroscopic scale [11]. At very small distances near the Planck length,
modifications have included discrete spacetime, breaking of discrete symmetries, Lorentz symmetry violation,
nonlocal interaction, extra dimensions, and non-commutative coordinates. However, these modifications are extremely
tiny to be noticed at their current observational levels. Decoupling at the Planck scale prevents these effects from
being observed at low energies. At large distances, modifications have included varying speed of light, varying
gravitational constant, modifying Newton's Second Law of motion, non-symmetric metric and incorporating scalar,
vector, and tensor particles into Einstein's gravity. The difficulty at this end is to obtain agreement with all
astrophysical and solar system observations. So far none of these alternative theories of gravity has succeeded in
replacing general relativity as the best theory of gravity.\\

A more fruitful approach to understand gravity is to develop quantum field theory of particles in curved spacetime [12].
This is done by treating spacetime classically and matter fields quantum mechanically. It is possible to study particle
creation in strong gravitational fields. This has led to the prediction of Hawking radiation in which particles are
emitted from a black hole with a thermal spectrum [13]; the Unruh effect in which an observer under acceleration 
in vacuum sees a thermal collection of particles [14]; and interpreting Einstein's equation as a thermodynamic
equation of state of spacetime and matter [15], thereby realizing toward an emergent theory of gravity [16].\\

\vspace{.2in}
\noindent
{\bf 6 \hspace{.1in} Is Spacetime Quantum?}\\

\noindent
In special relativity, the Lorentz transformation is a pseudo-rotation in four-dimensional Minkowski spacetime.
It is not possible to include the Planck constant or any other parameter into the transformation. It is a purely
mathematical transformation. Therefore there is no such theory to be called quantum theory of special relativity.
This term has a completely different meaning from relativistic quantum mechanics which is a description of matter.
Similarly, it is not possible to include Planck's constant in general coordinate transformations, or to have a
quantum theory of Riemannian spacetime. Since geometry is gravity in general relativity, this calls into the question
whether gravity really needs to be quantized [17][18]. Spacetime originally is a macroscopic concept. Is it possible
that Einstein's equation is similar in nature to Navier-Stokes equation in fluid mechanics as a macroscopic theory [19]?\\

The investigation of quantum black holes [20] shows that they are extremely microscopic objects with a macroscopic mass.
Their Schwarzschild radius is equal to their Compton wavelength. They exist at the boundary between classical and
quantum regions. They obey the Laws of Thermodynamics and they decay into elementary particles. A quantum black hole
of the size of the Planck length $1.6 \times 10^{-33}$ cm has a mass of $2.2 \times 10^{-5}$ gm. Like the nucleus of
a heavy atom, quantum black holes may require the use of quantum mechanics but not necessarily quantum field theory
for their description. The difference between quantum mechanics and quantum field theory is tremendous - it is the
creation and annihilation of particles. There are no anti-black holes in general relativity. Therefore there are
no virtual pair creations and annihilations of black holes as in ordinary particles. Two black holes combine to form
another black hole according to the area non-decrease theorem. The resulting black hole evaporates according to
Hawking's description with a temperature. Quantum black holes are intrinsically semi-classical objects.\\

\vspace{.2in}
\noindent
{\bf 7 \hspace{.1in} Quantum Gravity In Crisis}\\

\noindent
An important result in cosmology was obtained recently which can elucidate the nature of spacetime down to the 
smallest scale. This is the observation of the highest energy gamma rays from a gamma ray burst GRB 090510 by the
Fermi Gamma-Ray Space Telescope [21]. A single $31$-GeV photon was detected from a source at a redshift of $z = 0.903$
which corresponds to a distance of $7.3$ billion light years from Earth. It was the last of the seven pulses in a
short burst that lasted for $0.829$ s. One of the two postulates of Einstein's special relativity is Lorentz invariance
in that all observers measure exactly the same speed of light in vacuum, independent of the motion of the source and
of the photon energy. In certain quantum theories of gravity, there is great interest in the possibility that
Lorentz invariance might be broken near the Planck scale due to quantum fluctuation of spacetime and the notion of
spacetime foam. A variation of photon speed is an indication that Lorentz invariance is violated. This may be revealed
by observing the sharp features in the gamma ray burst light-curves. If the spread in travel time of less than $0.9$ s
between the highest and lowest-energy gamma rays in the burst GRB 090510 is all attributed to quantum effects, then a
thorough analysis shows that any quantum effects in which the speed is linearly proportional to energy do not show up
until the distance is down to about $0.8 L_{Pl}$, which is below the Planck length. This result therefore rules out
a number of quantum gravity models that predict such linear variation with energy.\\

The gamma ray burst reported above is significant in that it allows for the exploration of spacetime near Planck length
by using effects accumulated over cosmological distances since direct access to Planck energy in experiments is not
possible. The result indicates that there is no evidence so far of any quantum nature of spacetime above the Planck
length. Spacetime there is smooth and continuous. The speed of light is constant and special relativity is right.
At the Planck length, quantum black holes would appear in observation and they effectively provide a natural cutoff
to spacetime. For observable purpose, it is not necessary to consider theories below the Planck length. Further
detections using gamma ray bursts with even higher energy photons will settle the question of quantum spacetime
definitively. It would be amazing that in effect spacetime is classical and there is no need for a quantum theory
of gravity. There would be an underlying theory for gravity which is not gravity, just as statistical mechanics
is the underlying theory of thermodynamics. Unification would have a very different meaning from the current
understanding involving quantum gravity as a fundamental premise.\\

\end{document}